**Title**

# Using sequence data to study spatial scales of interactions driving spread of Highly Pathogenic Avian Influenza in Great Britain

**Authors:**

Anna Gamża*[1], Sam Lycett[1], Aeron Sanchez[1], Rowland Kao[1,2]

1) The Roslin Institute, University of Edinburgh, Edinburgh, UK
2) School of Physics and Astronomy, University of Edinburgh, Edinburgh, UK

## Abstract

H5N1 highly pathogenic avian influenza (HPAI) has been recently circulating in previously unseen patterns. As the underlying causes are uncertain, we need a better understanding of the drivers of virus circulation, as they underpin the spread, and inform about the impact of HPAI. We combine infection and bird population data, using Random Forest models to quantify the relationship between spatially aggregated values of possible infection risk factors and viral phylogenetic data from HPAI outbreaks to define scales of interactions that influence virus circulation. As genetic proximity is correlated to transmission proximity, the genetic relationships of sequences and relevant variables indicate the scale and resolution of interactions driving transmission. We used H5N1 Sequences from Great Britain divided into two periods: Dec 2020-May 2022, and Jun 2022-Oct 2024. Final model with selected spatial scales shows that for the first period various wild bird abundance variables were the strong predictors of genetic distance, and for the second period, these were geographical distance, and collection date. Variables related to infections have different spatial scales for analysed periods, while variables describing the environment have more consistent scales. The differences in predictors for the periods bears further investigation, however may be related to differences in the virus, areas of spread, or seasonal factors. This is the first phylogenetic study indicating possible role for game birds in the circulation of HPAI and supports experimental studies but must be viewed with caution as our study does not indicate direct causation. We show that the scale of spatial aggregation is variable dependent and reflects the scale of underlying processes. In the future our findings may be used to study dynamics of HPAI and design targeted surveillance and intervention strategies.

## Introduction

H5N1 highly pathogenic avian influenza (HPAI) virus has been increasingly spreading worldwide since 2020, infecting wild birds and mammals with unprecedented levels of mortality, and threatening domesticated poultry production in many countries (Koopmans et al., 2024). Concerns over the continued transmission of H5N1 have recently increased with reports of extensive circulation in the US cattle population (CDC, 2024a; Neumann & Kawaoka, 2024) and recent reports of human cases (CDC, 2024b). In much of Europe, HPAI appears to have become at least temporarily endemic, i.e. not dependent on seasonal introductions by migrating birds (Fusaro et al., 2024; Sacristán et al., 2024). While there are microbiological clues about the increased virus fitness (Fusaro et al., 2024), studies are still ongoing to explain how the emergence of the new strain affects the whole host-pathogen-environment system (Sacristán et al., 2024). Understanding of the H5N1 transmission dynamic in context of the whole epidemiological system is crucial to assess risk and impact of the virus on both avian and mammal populations either kept on farms or in the wild, and to develop efficient intervention strategies aimed at reducing the risks to animals and humans and stopping infection spread.

In a pathogen-host-environment system as complicated as Avian Influenza (Bonilla-Aldana et al., 2024; Koopmans et al., 2024), defining and studying all types of important potentially infectious interactions is challenging. The proper definition of the system involves understanding of the biology of virus itself, and the local environment in which transmission occurs. However, understanding of the latter can be difficult, at least in part because data on infection in wildlife can be both fragmented and heavily biased (Barroso et al., 2024). The availability of pathogen phylogenetic data, however, substantially improves our ability to assess these environmental characteristics. Particularly, evolutionary phylogeographic approaches (Dellicour et al., 2016) are capable of estimating the impact of spatially linked environmental information into estimates of rates of geographical spread. Yet, identifying the spatial scale at which environmental factors are important remains challenging as does question of whether it is the spatial units defined by human activity, such as administrative units, or natural factors that matters the most.

Here, we propose a different approach, that directly exploits the relationships between genetic distance between the virus sequences, as this is highly correlated with the transmission distance between infected individuals (informing how closely two cases are connected on transmission tree), and in turn, transmission distance is correlated to the length of the transmission chains that are indicative of transmission risks (Gamża et al., 2023). Thus, for example, genetic distance between H5N1 samples may increase with geographical distance. However, if there are more Anseriformes (e.g. ducks, geese or swans) present and genetic distance does not significantly increase with distance, then Anseriforme density may decrease the length of transmission chains between spatially separated places, i.e. due to migration patterns that decouple genetic distance from spatial distance. On the other hand, if Anseriforme density indicates increase in genetic distance, this means that direct links between samples are less likely – and is thus indicative of the presence of missing factors impacting viral circulation of the virus. This may be because of unobserved circulation in the Anseriforme population but also may be due to other correlated factors. In addition, by considering separately genetic changes most closely linked to transmission with those most closely associated with co-infection and reassortment, we can identify ecological patterns associated with different kinds of evolutionary change (e.g. Harvey, Pinto et al. 2025).

Past studies on risk factors of Avian Flu, only considered risks mapped with predetermined, fixed resolution, e.g. see (Gargallo et al., 2022; Gilbert & Pfeiffer, 2012). However, identifying the scales of various interactions that are at play in transmission system such as HPAI can be challenging given the mobility of human and animal populations, and may also be variable, for example due to the characteristics of the epidemiology, or seasonal changes in causative factors, such as the environmental survival of virus (Stenkamp-Strahm et al., 2020). Assessing the risk at inappropriate scales (both too small and too large) can lead to un- or overestimation of the risks (Brock et al., 2019; Meentemeyer et al., 2012).

We consider recent data (December 2020 to May 2022, and from June 2022 to October 2024) from the epidemic of HPAI H5N1 in Great Britain and using Random Forest models, we ask how various spatially aggregated demographic and environmental variables impacts the genetic distance between sampled viruses. Crucially we ask what scales of aggregation are most informative and consider the performance of simple gridded aggregation against aggregation according to administrative units is more relevant, as this is an indicator of whether human, or natural activity is an underlying driver. We use the selected scales to fit the final Random Forest models that indicate how selected variables are indicative of the genetic distance relationships.

## Results

### *Scales of areas of aggregation*

We used Random Forest modelling and permutation importance scores to select scales of variables that can be aggregated over space, namely number of domestic bird HPAI H5N1 cases, number of wild bird HPAI H5N1 cases, poultry farm count and abundance of wild bird species (separately for 17 species groups). To select the scale, we assumed that the best scale is the one that on average provides aggregated variable that is the most important for the genetic distance determination (with highest importance score) and results in the best model fit (highest $R^2$)

The selected scales are presented in Tables 1 and 2. Out of three used metrices the $R^2$ was the most variable, while importance based metrices were more consistent. Since $R^2$ describes the fit of the whole model rather than fit to one variable that is changed, generally the differences of $R^2$ from our RF models are quite small. Moreover, as for abundance variables we changed the scales of all abundance species group variables together, the $R^2$ change does not indicate the goodness of fit when only one of the abundance variables is changed.

Table 1. Scale selected for variables with aggregated data based on Random Forest model fits relating genetic distance among viruses to ecological and epidemiological predictors for non-reassorting genome segments (HA, NA and M) as indicated by three matrices; in parentheses is the number of Random Forest model fit repetitions that selected the scale (out of 100) as the most informative one; underlined is the scale selected for the final model

| **Variable** | **Scale selected for season 1** | | | **Scale selected for season 2** | | |
|---|---|---|---|---|---|---|
| | Metrics I[1] | Metrics II[2] | Metrics III[3] | Metrics I[1] | Metrics II[2] | Metrics III[3] |
| **Number of domestic bird cases** | LA (33/100) | LA (84/100) | LA (95/100) | XL (62/100) | XL (77/100) | XL (83/100) |
| **Number of wild bird cases** | XL (50/100) | XL (96/100) | XL (91/100) | XXL (89/100) | XXL (100/100) | XXL (87/100) |
| **Poultry farm count** | S (86/100) | S (96/100) | S (97/100) | IZ (31/100) | L (49/100) | L (68/100) |
| **Abundance of corvids** | NA* | XL (96/100) | XL (91/100) | NA* | M (97/100) | M (74/100) |
| **Abundance of dabbling ducks** | NA* | L (66/100) | L (61/100) | NA* | M (99/100) | M (100/100) |
| **Abundance of diving ducks** | NA* | M (77/100) | XL (60/100) | NA* | M (99/100) | M (100/100) |
| **Abundance of game birds** | NA* | M (51/100) | XL (100/100) | NA* | M (100/100) | M (100/100) |
| **Abundance of geese** | NA* | L (90/100) | L (91/100) | NA* | S (100/100) | S (100/100) |
| **Abundance of gulls** | NA* | M (100/100) | S (99/100) | NA* | M (100/100) | M (100/100) |
| **Abundance of herons and egrets** | NA* | L (100/100) | L (100/100) | NA* | XL (51/100) | L (66/100) |
| **Abundance of near passerines** | NA* | M (84/100) | M (73/100) | NA* | L (46/100) | L (73/100) |
| **Abundance of owls** | NA* | L (98/100) | L (100/100) | NA* | S (89/100) | L (98/100) |
| **Abundance of passerines** | NA* | L (100/100) | L (100/100) | NA* | M (100/100) | M (100/100) |
| **Abundance of pigeons** | NA* | M (89/100) | M (80/100) | NA* | M (88/100) | M (97/100) |
| **Abundance of rails and grebes** | NA* | L (98/100) | L (94/100) | NA* | XL (50/100) | L (87/100) |
| **Abundance of raptors** | NA* | L (67/100) | L (100/100) | NA* | M (88/100) | M (73/100) |
| **Abundance of seabirds** | NA* | M (100/100) | M (84/100) | NA* | M (100/100) | M (100/100) |
| **Abundance of swans** | NA* | M (100/100) | M (100/100) | NA* | M (100/100) | M (100/100) |
| **Abundance of terns** | NA* | M (94/100) | XL (61/100) | NA* | M (100/100) | M (100/100) |
| **Abundance of waders** | NA* | M (91/100) | L (83/100) | NA* | M (79/100) | M (78/100) |
| **Abundance of All birds** | NA* | M (99/100) | L (100/100) | NA* | M (100/100) | M (100/100) |

1) Difference in $R^2$ between the model including variables in all scales and separate models each with tested variable in only one scale (tested scale)
2) Difference in permutation importance of the scaled variables from the model including variables in all scales
3) Difference in permutation importance of the scaled variables from separate models each with tested variable in only one scale (tested scale)

* $R^2$ change does not indicate the fit for each abundance variable separately, as the scale was changed for all abundance variables together

Table 2. Scale selected for variables with aggregated data based on Random Forest model fits relating genetic distance among viruses to ecological and epidemiological predictors for non-reassorting genome segments (HA, NA and M) across reassortant types as indicated by three matrices; in parentheses is the number of Random Forest model fit repetitions that selected the scale (out of 100) as the most informative one; underlined is the scale selected for the final model

| **Variable** | **Scale selected for season 1** | | | **Scale selected for season 2** | | |
|---|---|---|---|---|---|---|
| | Metrics I[1] | Metrics II[2] | Metrics III[3] | Metrics I[1] | Metrics II[2] | Metrics III[3] |
| **Number of domestic bird cases** | DZ (17/100) | <u>L</u> (95/100) | <u>L</u> (100/100) | <u>S</u> (26/100) | <u>XXL</u> (100/100) | <u>XXL</u> (100/100) |
| **Number of wild bird cases** | LA (21/100) | <u>XL</u> (99/100) | <u>XL</u> (99/100) | M (25/100) | <u>XXL</u> (100/100) | <u>XXL</u> (100/100) |
| **Poultry farm count** | S (97/100) | <u>M</u> (76/100) | <u>M</u> (62/100) | M (43/100) | <u>XL</u> (95/100) | <u>XL</u> (92/100) |
| **Abundance of corvids** | NA* | <u>M</u> (82/100) | <u>M</u> (90/100) | NA* | <u>L</u> (48/100) | <u>M</u> (83/100) |
| **Abundance of dabbling ducks** | NA* | <u>S</u> (87/100) | <u>S</u> (88/100) | NA* | <u>M</u> (91/100) | <u>M</u> (94/100) |
| **Abundance of diving ducks** | NA* | <u>M</u> (89/100) | <u>M</u> (92/100) | NA* | <u>S</u> (100/100) | <u>S</u> (100/100) |
| **Abundance of game birds** | NA* | S (64/100) | <u>XL</u> (82/100) | NA* | M (46/100) | <u>XXL</u> (86/100) |
| **Abundance of geese** | NA* | <u>S</u> (90/100) | <u>S</u> (91/100) | NA* | <u>M</u> (96/100) | <u>M</u> (98/100) |
| **Abundance of gulls** | NA* | <u>S</u> (88/100) | <u>S</u> (100/100) | NA* | <u>M</u> (100/100) | <u>M</u> (99/100) |
| **Abundance of herons and egrets** | NA* | <u>M</u> (89/100) | <u>M</u> (46/100) | NA* | <u>L</u> (51/100) | <u>L</u> (66/100) |
| **Abundance of near passerines** | NA* | <u>M</u> (99/100) | <u>M</u> (96/100) | NA* | S (45/100) | <u>XL</u> (70/100) |
| **Abundance of owls** | NA* | <u>M</u> (96/100) | <u>M</u> (65/100) | NA* | S (73/100) | <u>XL</u> (95/100) |
| **Abundance of passerines** | NA* | <u>L</u> (88/100) | <u>L</u> (100/100) | NA* | M (86/100) | <u>S</u> (99/100) |
| **Abundance of pigeons** | NA* | <u>M</u> (100/100) | <u>M</u> (100/100) | NA* | <u>M</u> (75/100) | <u>M</u> (79/100) |
| **Abundance of rails and grebes** | NA* | L (60/100) | <u>XL</u> (67/100) | NA* | M (51/100) | <u>XL</u> (98/100) |
| **Abundance of raptors** | NA* | <u>XL</u> (79/100) | <u>XL</u> (96/100) | NA* | <u>L</u> (95/100) | <u>L</u> (100/100) |
| **Abundance of seabirds** | NA* | <u>S</u> (80/100) | <u>S</u> (96/100) | NA* | <u>M</u> (100/100) | <u>M</u> (100/100) |
| **Abundance of swans** | NA* | <u>M</u> (100/100) | <u>M</u> (100/100) | NA* | <u>M</u> (100/100) | <u>M</u> (100/100) |
| **Abundance of terns** | NA* | <u>M</u> (80/100) | <u>XL</u> (100/100) | NA* | <u>M</u> (88/100) | <u>M</u> (91/100) |
| **Abundance of waders** | NA* | <u>M</u> (67/100) | <u>M</u> (52/100) | NA* | <u>S</u> (53/100) | <u>S</u> (48/100) |
| **Abundance of All birds** | NA* | M (100/100) | M (100/100) | NA* | M (100/100) | M (100/100) |

1) Difference in $R^2$ between the model including variables in all scales and separate models each with tested variable in only one scale (tested scale)
2) Difference in permutation importance of the scaled variables from the model including variables in all scales
3) Difference in permutation importance of the scaled variables from separate models each with tested variable in only one scale (tested scale)

* $R^2$ change does not indicate the fit for each abundance variable separately, as the scale was changed for all abundance variables together

Maps illustrating areas in selected scales and shapes are presented on Supplementary Figures 7-14.

*Final model*

We used variables aggregated over chosen scales (Table 1 and 2) to generate the final fits of the Random Forest models. The importance plot for the final model for each of the two periods are presented on Figures 1 and 2 together with the Shapley Additive exPlanations (SHAP) values describing the dependence between genetic distance and the most important variables for each variable separately (Lundberg & Lee, 2017). In figure 1, we compare genetic disatcnes based on non-reassorting segments of the virus only (HA, NA and M), considering all pairs of available samples reported within 30 days of each other, with sequence 1 representing the earlier sample. In figure 2, we also use non-reassorting genome based genetic distance, but only compare sequences across reassortant types, with sequence 1 representing the sample coming from an earlier reassortant. Comparing two Avian Flu seasons, there are clear differences not only in scales of variables accumulation but also in the relative importance of the variables.

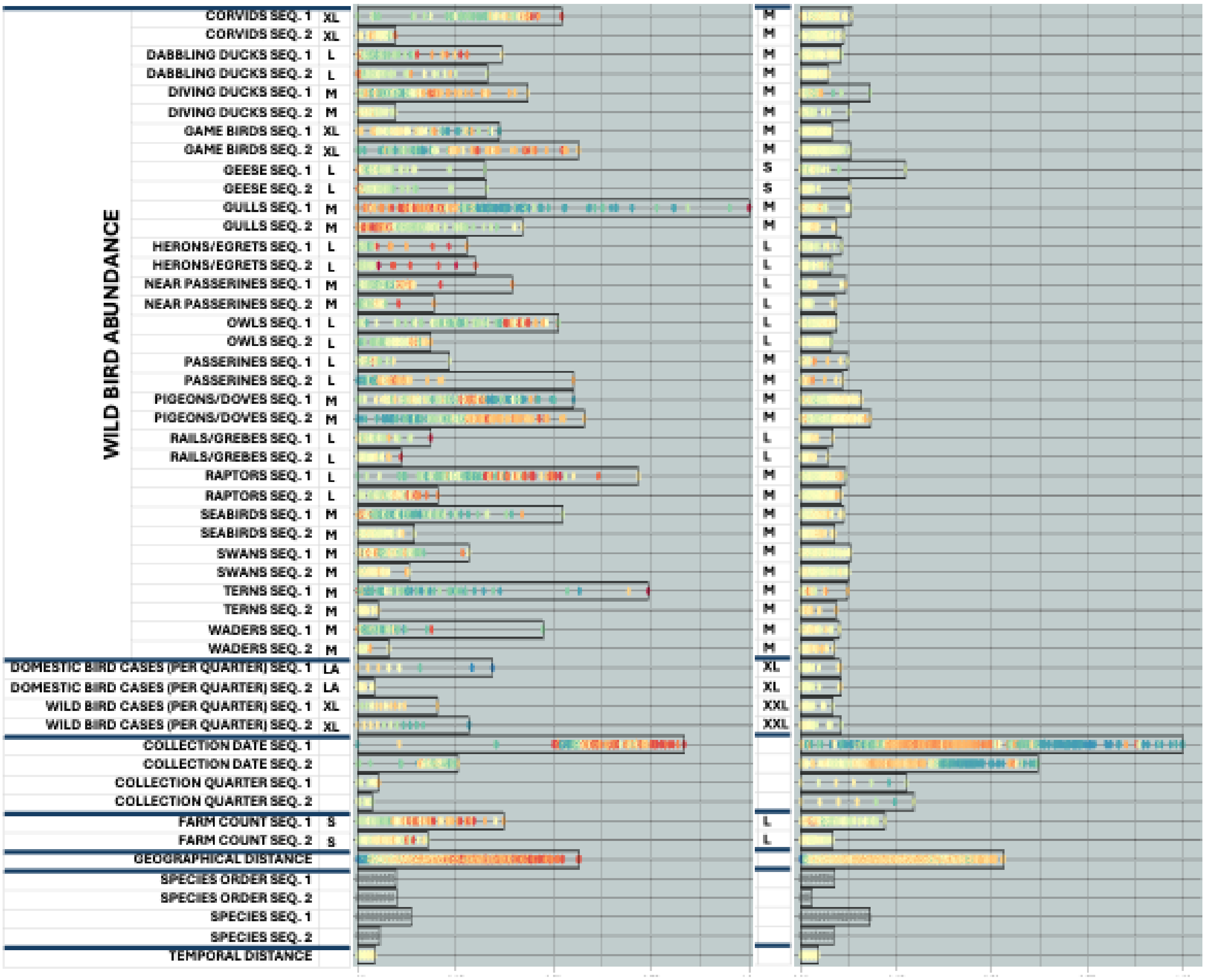

Figure 1. Random forest models relating genetic distance among viruses to ecological and epidemiological predictors for non-reassorting genome segments (HA, NA and M) during Period I (December 2020-May 2022; left) and Period II (June 2022-October 2024; right). Horizontal bars show global SHAP importance, indicating the relative contribution of each predictor to model predictions across the dataset. Predictors are shown at the most relevant spatial scales, from S (3 × 3 km grid cells) to XXL (108 × 108 km grid cells); LA denotes local authority level (Table 1). Coloured points show SHAP values for individual observations as a function of predictor values. Positive SHAP values (red) indicate that a predictor contributes to higher predicted genetic distance, whereas negative SHAP values (blue) indicate a contribution to lower predicted genetic distance. The concentration of points reflects the distribution of observations across the range of predictor values. Grey bars denote categorical predictors.

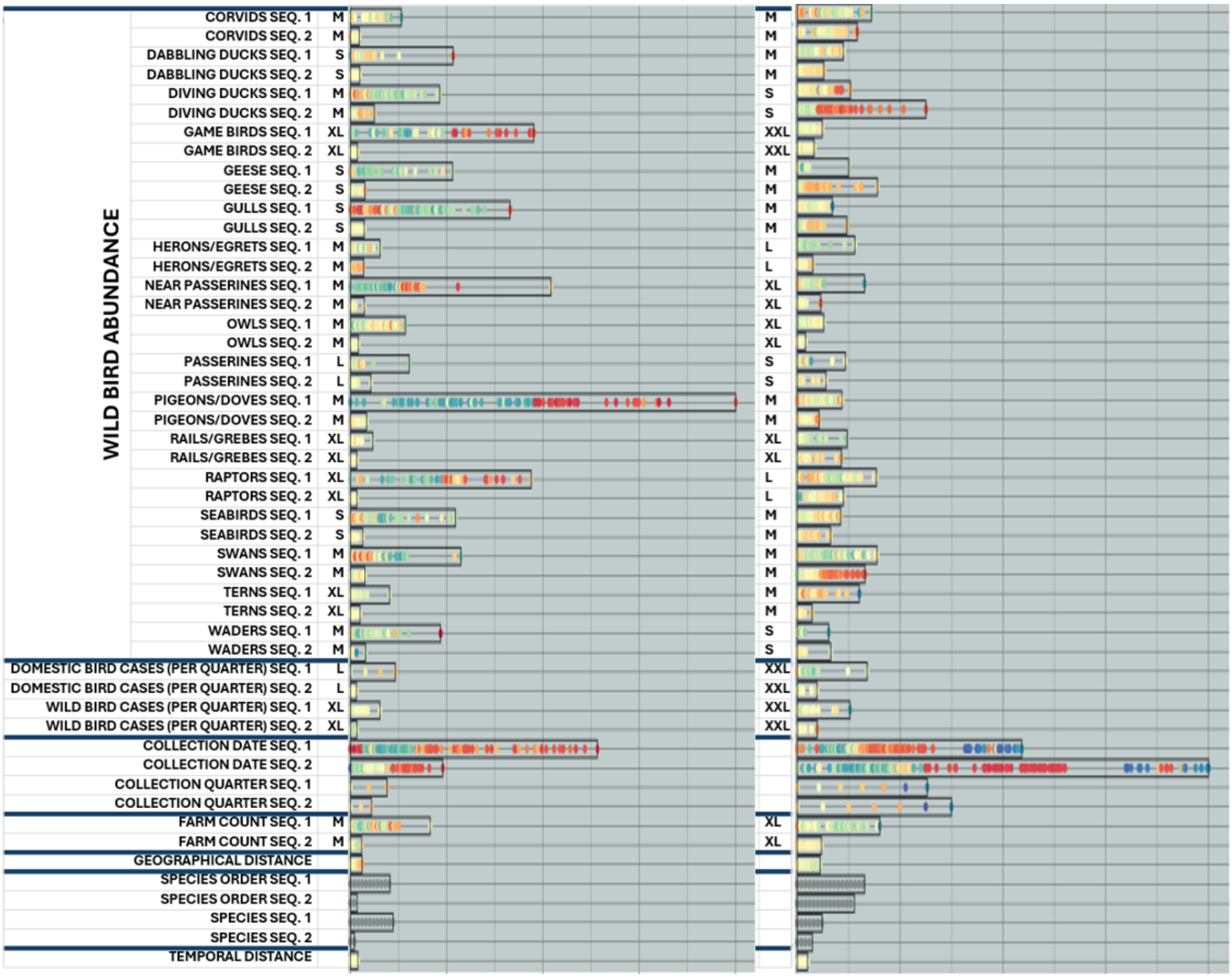

Figure 2. Random forest models relating genetic distance among viruses to ecological and epidemiological predictors for non-reassorting genome segments (HA, NA and M) across reassortant types, during Period I (December 2020-May 2022; left) and Period II (June 2022-October 2024; right). Horizontal bars show global SHAP importance, indicating the relative contribution of each predictor to model predictions across the dataset. Predictors are shown at the most relevant spatial scales, from S (3 × 3 km grid cells) to XXL (108 × 108 km grid cells). Coloured points show SHAP values for individual observations as a function of predictor values. Positive SHAP values (red) indicate that a predictor contributes to higher predicted genetic distance, whereas negative SHAP values (blue) indicate a contribution to lower predicted genetic distance. The concentration of points reflects the distribution of observations across the range of predictor values. Grey bars denote categorical predictors.

## Discussion

Here, we have used genetic distance between sequences of H5N1 HPAI in Great Britain to identify which environmental and epidemiological factors are the most informative in predicting those genetic distances separately for two periods Dec 2020 to May 2022 and June 2022 to Oct 2024. As these genetic distances are indicating virus transmission, the most informative factors are likely to be important indicators of factors driving (possibly unobserved) virus circulation.

The scales selected for variables that were not directly dependent on the influenza outbreaks, namely the abundance of wild birds were largely consistent between periods. The abundance of all wild bird species when grouped together with a medium resolution (9km x 9km) was selected for both periods. There was however a difference in spatial scales for specific groups of birds.

Considering only the most important identified variables, the use of SHAP plots (extracting the univariate relationships between identified risk factors and genetic distances once correlated effects are removed) provides further clues as to possible causes of the observed patterns. Interpretation of large discontinuities in the SHAP values must be viewed with caution as, given the size of the dataset, they may be due to single anomalies in the data relationships. However more consistent SHAP trends can be informative. For example, the increase of genetic distance with geographical distance is consistent with our intuitive sense that distance at a large enough scale, is likely to decrease the probability of a direct contact between sample infections. The role of collection date in period 2 may reflect seasonal changes, with the large genetic distances over the summer period, reflecting the few infections are observed (in previous years, typically none) as these are likely related to multiple small outbreaks driven by hidden infections in contrast to big outbreaks registered on farms in Autumn of 2022. Increased sampling over time especially for more easily detectable farm outbreaks could result in fewer 'missing gaps' in the data.

Other informative factors detected by our model however merit further discussion. In period 1 (December 2021 to May 2022) for the across reassortant model game bird (with pheasants by far the most prevalent gamebird species in GB), appear to play an important role in predicting genetic distances. Higher abundance is associated with greater genetic distance and thus missing links which may be due to these species having a bridging role between observed samples with significant undetected transmission among them. While the relationships identified here are correlations and may not be causative, additional evidence from experimental data shows that pheasants in particular are both highly susceptible to HPAI and shed virus in faeces with unusually long survival times in the environment (Seekings et al., 2024), moreover they have been the subject of speculation on a potential role in the circulation of H5N1 (Defra's Science Advisory Council, 2023). The decrease in the observed role of game birds during period 2 when these birds are largely housed and not in the wild is also indicative (Game & Wildlife Conservation Trust, 2024).

For the non-reassorting genome model areas of higher abundance of gulls indicate much shorter genetic connections, potentially indicating a direct role for these species in bridging otherwise unconnected patterns of circulation. However, such considerations are at best only indicators of future directions of investigation.

When comparing the two periods: from December 2020 to May 2022, and from June 2022 to October 2024 there were significant differences both in terms of scales selected and variables dependence with genetic distance for final model fit, indicating that drivers in the circulation of H5N1 may vary substantially. This was the case when both comparing across periods for non-reassorting segments alone (Fig. 1) and across reassortant types (Fig. 2). Some of these differences are substantial. Nonetheless, the variability of scale across risk factors, and time periods that we have identified illustrates the importance of considering of relevant spatial scale in understanding transmission in complex disease ecosystems, as found in other systems (Brock et al., 2019). Overall, the methodology we developed can be used to study geographical scales of complex pathogen-host-environmental systems like avian influenza combining pathogen sequence data with many variables reported in spatial resolution. In the future the investigations of scales can be implemented to improve spatial sampling strategies, define targeted interventions and the zones they are implemented in, or create more accurate risk maps.

## Methods

### *Data*

For the analysis we integrated HPAI H5N1 sequences from period 03$^{rd}$ December 2020 to 19$^{th}$ October 2022 with all positive HPAI H5N1 cases from the same period, poultry farm data as of December 2022 and wild bird abundance data from eBird platform from year 2021; all data sources are summed up in Table 3.

Table 3. Raw data used for analysis of dependence of genetic distance between sequences of H5N1 and other variables connected to HPAI epidemic in years 2020-2022

| Dataset | Source | Description | Notes |
|---|---|---|---|
| **HPAI Sequences** | GISAID (GISAID Data Science Initiative) | Genetic sequence of HPAI cases from United Kingdom from years 2020-2022 | The coordinates attributed for each published sequence we analysed were provided by Animal and Plant Health Agency. |
| **HPAI positive cases in birds** | World Animal Health Information System (World Organisation for Animal Health, 2024) | Locations and metadata of HPAI outbreaks reported to World Organisation of Animal Health in years 2020-2022 | Both outbreaks on poultry farms as well as in wild birds or other bird premises were downloaded. |
| **Poultry farm count** | Great Britain Poultry Register (GBPR) | Extract from GBPR as of 1 December 2022 with poultry premises locations. | |
| **Wild bird abundance data** | eBird (Cornell Lab of Ornithology) | Weekly counts from 2021 in medium and low resolution. | |
| **Census geography shapefiles** | UK Data Service (UK Data Service) | Shapefiles for administrative areas (Output Areas) defined for 2011 Census | |
| **Postcode lookup with administrative areas codes** | Office for National Statistics (Office for National Statistics) | A best-fit lookup between postcodes, 2011 Census Output Areas, Lower Layer Super Output Areas/Data Zones, Middle Layer Output Areas/Intermediate Zones and Local Authority Districts | Lower and Middle Layer Super Output Areas are areas defined for England and Wales (Office for National Statistics), while Data Zones and Intermediate zones are areas defined for Scotland (Scotland's Census, 2021) |

We divided the sequence data into two periods that we analysed separately: first period spanning from 03rd December 2020 to 31st May 2022 and second from 01st June 2022 to 19th October 2024. This way we compared the most recent HPAI H5N1 season to the more historical outbreaks.

*Areas*

Table 4. Map based areas of aggregation for study of scale of interaction for HPAI H5N1 transmission in birds in Great Britain.

| Area | Description | Shape | Size |
|---|---|---|---|
| **Grid areas Size H** | High resolution grid areas | Rectangular grid | grid size approx. 3 x 3 km |
| **Grid areas Size M** | Medium resolution grid areas | Rectangular grid | grid size approx. 9 x 9 km |
| **Grid areas Size L** | Low resolution grid areas | Rectangular grid | grid size approx. 27 x 27 km |
| **Grid areas Size XL** | Extra-Low resolution grid areas | Rectangular grid | grid size approx. 54 x 54 km |
| **Grid areas Size XXL** | Extra-Extra-Low resolution grid areas | Rectangular grid | grid size approx. 108 x 108 km |
| **Output area** | • Output Areas (OAs) are the lowest level of geographical area designed for 2011 census statistics, they are created by gathering together postcodes.<br>• A single output area contained at least 50 people and 20 households.<br>• Output areas form the building blocks for all other census geographies. | Different population densities mean the size and shape of output areas can vary greatly | • In England and Wales, contain between 40 and 250 households and population between 100 and 625 persons as of 2011<br>• In Scotland, contains at least 50 people and 20 households as of 2011 |
| **Lower layer Super Output Areas/Data Zone** | • Areas composed of groups of 2011 Census Output Areas, usually four or five, in England and Wales called Lower Layer Super Output Areas and in Scotland, called Data Zones | Different population densities mean the size and shape of output areas can vary greatly | • Lower Layer Super Output Areas contain between 400 and 1,200 households and population between 1,000 and 3,000 persons as of 2011<br>• Data Zones designed to have population of 500 to 1,000 persons as of 2011 |
| **Middle layer Super Output Areas/Intermediate zone** | • Areas composed of groups of Lower Layer Super Output Areas/Data Zones usually four or five, in England and Wales called Middle layer Super Output Areas and in Scotland, called Intermediate zones. They fit within local authorities. | Different population densities mean the size and shape of output areas can vary greatly | • Middle layer Super Output Areas contain between 2,000 and 6,000 households and population between 5,000 and 15,000 persons as of 2011<br>• Intermediate Zones designed to have population of 2,500 to 6,000 persons as of 2011 |
| **Local authority** | The Local Authority Area list contains 378 areas in Great Britain of the following constituent geographies:<br>• 36 Metropolitan Districts in England<br>• 201 Non-Metropolitan Districts in England<br>• 31 London Boroughs in England<br>• 54 Unitary Authorities in England<br>• 32 Council Areas in Scotland<br>• 22 Unitary Authorities in Wales<br>• 2 Census Merged Districts in England | Different population densities mean the size and shape of output areas can vary greatly | |

Output Area shape files were provided by the UK Data Service (UK Data Service, 2024); we used versions defined for 2011 census. Their shape and size are optimised to account for variable human population densities, such that each area houses approximately similar number of inhabitants (National Records of Scotland; Office for National Statistics, 2024a). As the coordinates were provided in British National Grid coordinates (BNG, ESPG:27700) we converted them in R (function: st_transform; package: sf (Edzer Pebesma, 2024)) into latitude and longitude format (lat/long, WGS84; EPSG:4326). Dataset with postcode lookup containing codes for 2011 Census Output Areas as well as codes for Lower Layer Super Output Areas/Data Zones, Middle Layer Super Output/Intermittent Zones and Local Authority Districts was downloaded from Office for National Statistics (Office for National Statistics, 2024b).

Grid areas of various sizes were based on the grids over which eBird wild bird abundance (Cornell Lab of Ornithology, 2024) data were aggregated. Bigger size grids (XL and XXL size) were created as multiples of M size grid by merging 6x6 areas of M resolution for XL scale and 12x12 areas of M resolution for XXL scale (functions: st_bbox and st_polygon; package: sf (Edzer Pebesma, 2024)).

Wild bird abundance data were aggregated over grid areas only (sizes from M to XXL). Poultry farm data were aggregated on grid areas (sizes from H to XXL) as well as administrative areas (Output Zones, Data Zones, Intermediate Zones and Local Authorities). Positive HPAI cases from domestic and wild birds were aggregated separately on grid areas (sizes from H to XXL) as well as administrative areas (Data Zones, Intermediate Zones and Local Authorities).

*Data processing*

Of HPAI virus sequences from the United Kingdom in the period from 30th December 2020 to 12th October 2024 obtained from GISAID and connected to coordinates, 778 were type H5N1 from Great Britain. Approximate coordinates of the location of sequenced cases were provided by Animal and Plant Health Agency. The coordinates were provided in WGS84 system and described with latitude and longitude format (lat/long, WGS84; EPSG:4326). Sequences were reported with species names; during the data processing obvious typographical errors were replaced with standardised species names. For the standardisation we assumed that each case reported without precise species information was representing the most common species in family, such that “duck” (2 cases) or “goose” (2 cases) (i.e. without any further specification) was a “domestic duck/goose”, “buzzard” (3 cases) was “common_ buzzard” and “falcon” (1 case) was “peregrine_falcon”.

Public data with reported HPAI H5N1 outbreaks (“Influenza A viruses of high pathogenicity (Inf. with)”; subtype: H5N1) from Great Britain (excluding reports from Isle of Man, Jersey, Alderney and Guernsey; and cases from Northern Ireland) from 3rd December 2020 to 9th August 2025were downloaded from World Animal Health Information System (WAHIS) system (World Organisation for Animal Health, 2024). We included both reports classified as “poultry” and “non-poultry including wild birds”; where the later included not only wild bird cases but also captive, domestic bird cases from hobby farm, zoos and similar premises. The locations of the outbreaks were exported from the WAHIS map and connected to metadata downloaded from outbreak list for each report using their Outbreak IDs. The approximate coordinates attached to cases were provided in WGS84 system and described with latitude and longitude format (lat/long, WGS84; EPSG:4326). Outbreak IDs from metadata were extracted from the reference description (containing reference code and Outbreak ID). We standardised reference codes such that they follow the same format: “AIV [year]/[number]” for domestic birds and “WB [number]” for wild bird cases. After matching procedures, 1,654 records were directly linked; a further seven were matched using date and location consistency, and four missing from the map (longitude = 0) were added manually. One unmatched mapped point was excluded. In total we included 1,210 wild cases and 444 domestic cases. The map of the included cases is provided in Supplementary Information (Supplementary Figures 1-6).

Next, to all the HPAI H5N1 cases we assigned grid areas (considering all resolutions from H to XXL, as defined in Table 3) and 2011 Census Output Areas (as defined in Table 3). To assign areas, for each point we found the areas with which the point coordinates overlap (function: st_intersects; package: sf (Edzer Pebesma, 2024)). If the point lay at the boundary between two areas, one of the two was randomly selected. Using the Output Zone assignment other administrative areas were assigned (as defined in Table 3) using a dataset with postcode lookup containing codes for 2011 Census Output Areas as well as codes for Lower Layer Super Output Areas/Data Zones, Middle Layer Super Output/Intermittent Zones and Local Authority Districts downloaded from Office for National Statistics (Office for National Statistics, 2024b). Additionally, we assigned country (England, Scotland or Wales) to each case using the shape file with country borders and assigning country that overlaps with each data point (function: st_intersects; package: sf (Edzer Pebesma, 2024)). When no country was assigned we generated small area (approx. 2.5 km) around each point (case coordinates) and searched for overlapping country borders, the areas was subsequently reduced 2 to 28 times until just one country border was assigned. If the country still was not found the procedure was repeated with the initial area of approx. 1 km. This was done to minimise effect of the uncertainty of approximate coordinates and country borders.

H5N1 case data were divided into wild bird data and domestic bird data based on outbreak category registered in WAHIS. For data from period from 3rd December 2020 to 9th August 2025, we included total of 1656 H5N1 cases of which 1210 were from wild birds and 444 from domestic birds. To each case the quarter was assigned based on the date when the outbreak started. The number of positive HPAI H5N1 cases for domestic and wild birds separately in Great Britain per quarter is presented in Table 4. Next, we aggregated separately domestic and wild bird cases both in space (over all areas) and time (over quarters) to obtain case counts for each area (using R package Data Table with natation: [ , .(“casesNo”=.N), by=c(<area type>, “Collection quarter”).

*Table 5. Total number of wild bird and domestic bird cases reported to WAHIS in each quarter from period 3rd December 2020 to 9th August 2025*

| Start quarter | No. wild bird cases | No. domestic bird cases |
|---|---|---|
| **2020/4** | 4 | 1 |
| **2021/1** | 2 | 1 |
| **2021/2** | 1 | 0 |
| **2021/3** | 3 | 0 |
| **2021/4** | 164 | 77 |
| **2022/1** | 108 | 34 |
| **2022/2** | 72 | 11 |
| **2022/3** | 81 | 45 |
| **2022/4** | 219 | 144 |
| **2023/1** | 62 | 21 |
| **2023/2** | 67 | 10 |
| **2023/3** | 86 | 18 |
| **2023/4** | 15 | 5 |
| **2024/1** | 2 | 1 |
| **2024/2** | 1 | 0 |
| **2024/3** | 0 | 0 |
| **2024/4** | 20 | 14 |
| **2025/1** | 214 | 45 |
| **2025/2** | 58 | 8 |
| **2025/3** | 31 | 9 |

Public wild bird abundance data describing the abundance of each species for every week of year 2021 were downloaded from eBird platform. Relative abundance distributions were retrieved from eBird Status and Trends using R package ebirdst (Matthew Strimas-Mackey, 2022); the details of the relative abundance generation can be found in (Vickers et al., 2024). The estimates were provided for the full year 2021 at weekly intervals, across a regular grid. We used grids of 2 sizes (M and L in Table 4). Relative abundance distributions for Great Britain were available for 256 species, with seasonal quality ratings assessed between 0 and 3. We consider only those species with scores of 2 or 3 during the modelled period (to exclude predictions with low confidence) and a sum of relative weighted abundance scores within the relevant time period of >1 (to exclude uncommon and rare species). This produced a final pool of 152 species that were next aggregated into 9 species groups: corvids, ducks, game birds, gulls, owls, raptors, swans, terns, and waders. For each group we calculated mean abundance averaging over all weekly reported abundances.

Protected data on farm locations as of 1st December 2022 were downloaded from Great Britain Poultry Register (GBPR). In total the data described 58 119 poultry premises. As the farm coordinates were provided in British National Grid, described with Easting and Northing coordinates (BNG, ESPG:27700) we converted them in R (function: st_transform; package: sf (Edzer Pebesma, 2024)) into latitude and longitude format (lat/long, WGS84; EPSG:4326). To calculate the farm count for all areas, first for each farm we found the areas (grid areas of sizes H to XXL as well as 2011 Census Output Areas) with which the farm's point coordinates overlap (function: st_intersects; package: sf (Edzer Pebesma, 2024)). If the point was lying on the border, one area was randomly selected. Using the Output Zone assignment other administrative areas were assigned (as defined in Table 3) using a dataset with postcode lookup containing codes for 2011 Census Output Areas as well as codes for Lower Layer Super Output Areas/Data Zones, Middle Layer Super Output/Intermittent Zones and Local Authority Districts downloaded from Office for National Statistics (Office for National Statistics, 2024b). Next, we aggregated the data to obtain farm count for each area (R function: table applied to the column with assigned area of aggregation).

To combine H5N1 sequence data with wild bird case data, domestic birds case data, wild abundance data and farm count, we assigned to each sequence the aggregated number of cases on their quarter and assigned areas of various shapes and sizes (as defined in Table 4), and mean wild bird abundance and farm count aggregated on areas only. For the across reassortant analysis, reassortments were classified via segment-specific clustering, performed using nearest-neighbour hierarchical clustering (bioseq package) with a 0.5% threshold for polymerase segments (PB2, PB1, PA) and 1% for others. Clusters were ranked by size and assigned numerical identifiers to generate an eight-segment barcode for each genome.

To study the relationship between genetic distance and aggregated variables we generated pairs between all sequenced cases to which metadata and aggregated data was assigned as described above.

Within each pair, sequences were ordered by inferred emergence timing. Genetic distances were calculated between HA, NA, and M segments and the earliest sequence from the reference strain (H5N1/2020/R1; EPI_ISL_603133). Reassortants were then ranked by mean distance to this reference (Table 6), with smaller distances indicating earlier emergence. For pairs from different reassortants (n = 181,521), Sequence 1 corresponded to the reassortant genetically closest to the earliest reference strain H5N1/2020/R1 (Harvey, 2025). For pairs from the same reassortant (n = 102,456), Sequence 1 was defined as the earlier sampled sequence. Where both sequences were sampled on the same date (n = 568), ordering was assigned at random.

*Table 6. Order of reassortants based on the mean distance between Ha-Na-M segments from analysed sequences and from the earliest H5N1/2020/R1 sequence.*

| Reassortant code* | Mean genetic distance to sequence | Standard deviation | No. sequences | Final order |
|---|---|---|---|---|
| **H5N1/2021/R3_Europe** | 0.0043 | 0.0002 | 2 | 1 |
| **H5N1/2020/R1_Europe** | 0.0087 | 0.0027 | 159 | 2 |
| **H5N1/2021/R2_Europe** | 0.0124 | 0.0021 | 399 | 3 |
| **H5N1/2022/R15_Europe** | 0.0140 | 0.0014 | 10 | 4 |
| **H5N1/2022/R10_Europe** | 0.0142 | 0.0015 | 177 | 5 |
| **H5N1/2022/R17_Europe** | 0.0163 | 0.0052 | 3 | 6 |
| **H5N1/2021/R5_Europe** | 0.0164 | 0.0003 | 7 | 7 |
| **H5N1/2023/R3_Europe** | 0.0166 | 0.0002 | 8 | 8 |
| **H5N1/2023/R9_Europe** | 0.0170 | NA | 1 | 9 |
| **H5N1/2021/R4_Europe** | 0.0183 | NA | 1 | 10 |

*See (Harvey et al., 2025) for naming convention

The genetic distance between paired sequences was calculated for each pair using the Tamura and Nei 1993 model (Tamura & Nei, 1993) (R; function: dist.dna, package: APE (Emmanuel Paradis, 2024)). Additionally, we calculated absolute geographical distance pair (R; function: st_distance; package: sf) and absolute temporal distance (module of the difference in days). All studied variables assigned to genetic distances are presented in Supplementary Table 1.

*Random Forest model*

To describe the relationship between genetic distance for paired sequenced cases and other variables we considered many versions of the Random Forest model for which we estimated $R^2$ and feature importance (permutation importance). We selected the final version presented on Fig. 1 and 2 in the Results section for which we generated Accumulated Local Effects (ALE) as well as SHAP plots to gain additional insight into selected variables.

Random Forest analysis was implemented in R using package "ranger" (Marvin N. Wright, 2023) that allowed us to perform multicore calculations that substantially reduce computing time. Each model run was performed with the same set of parameters. Number of trees was set at 500 and number of variables to possibly split at in each node (mtry) was set at square root of the number of variables. The output was generated for permutation importance. Other parameters regarding Random Forest fitting were set at default values as defined by the package authors. Each model version was run for all the data (i.e. 100% data used for training) to get the best insight into patterns that can be captured by the algorithm.

*Temporal distance optimization*

To exclude from the dataset pairs of sequences that differ from each other most likely because of the time of emergence we implemented a temporal distance cutoff excluding the sequences that were sampled over 30 days from each other. The 30-day threshold was selected based on temporal distance optimisation, where we compared importance of temporal distance and $R^2$ of the Random Forest model. The optimisation was done on all data (two periods analysed together), and aimed at selecting the model with temporal distance threshold that had the lowest importance (indicating the genetic variability within the threshold is less likely to be due to the time of emergence) and in the same time the biggest $R^2$ (to ensure the least amount of data is rejected). The results from the genetic distance threshold optimisation are presented on Fig. 3.

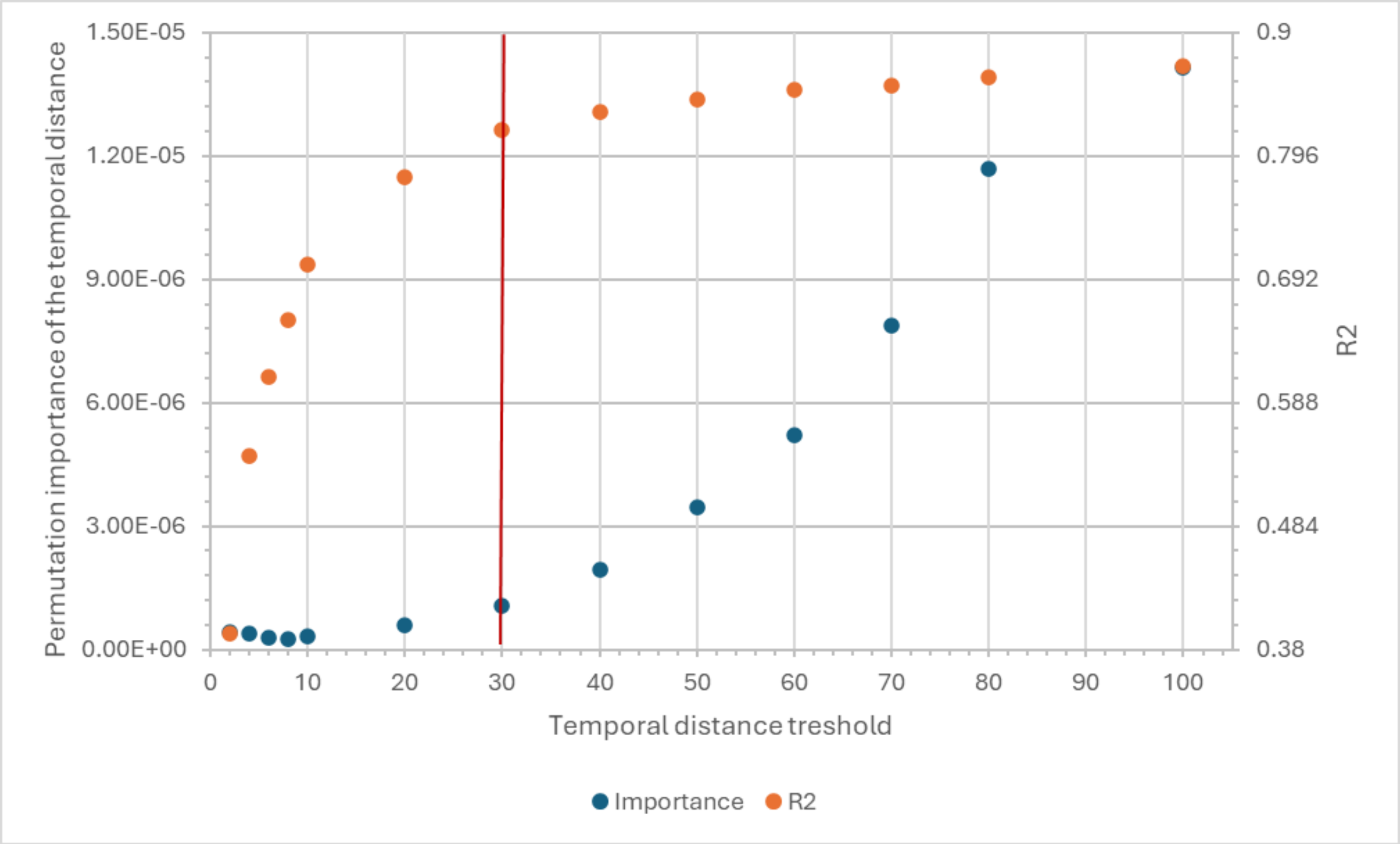

Figure 3. Temporal distance threshold optimisation implemented to select threshold for pairwise for Random Forest modelling that minimizes the importance of temporal distance (left axis) without compromising the model fit (indicted by $R^2$; right axis); select optimal threshold of 30 days marked with red line.

*Scale selection*

We selected the scales of spatially aggregated variables comparing RF runs with variables defined on different scales by using 3 metrices: 1) the difference in $R^2$ between model using all scales and model using just one scale; 2) permutation importance of the scale variables in RF model with all scale variables; 3) permutation importance of the scale variables in RF model with just this one variable. We repeated RF for each variable 100 times and selected the scale that was indicated for majority of runs. If there was the discrepancy in selected scale between metrices the final selected scale was the one for metric that selected the scale with the biggest score (total number of times particular scale was selected in 100 runs).

For the abundance variables we compared model containing variables aggregated on all scales included for all bird species and models containing all 17 bird group abundance variables in one (the same) spatial scale. We did not generate separate RF scale versions for each abundance variable (keeping all the other variables unchanged) to minimise the number of RF runs, therefore first metrics (based on the model fit difference) does not select for each abundance group separately.

The final model was constructed using variables describing HPAI H5N1 case aggregates, farm counts and wild abundance calculated over selected scales (see Table 1 and 2). For all numeric variables Shapley (SHAP) values were calculated using the fastshap package (100 simulations, regression-based adjustment) (Lundberg & Lee, 2017; Greenwell, 2026). Random Forest models were trained on the 100% data and keeping number of intervals (“k”) set to be 500 (or the maximum number of intervals if less than 500 data points available).

**Acknowledgments**

This study was funded by Strategic Program grant to Roslin Institute (grant no. BB/P013740/1), Flu-MAP project (grant no. BB/X006123/1), Flu-TrailMap (grant number BB/Y007271/1, BB/Y007298/1) and FluTrailMap- One Health (MR/Y03368X/1), as well as it was supported by the Scottish Government Rural and Environment Science and Analytical Services Division as part of the Center of Expertise on Animal Disease Outbreaks (EPIC).

**Supplementary Information**

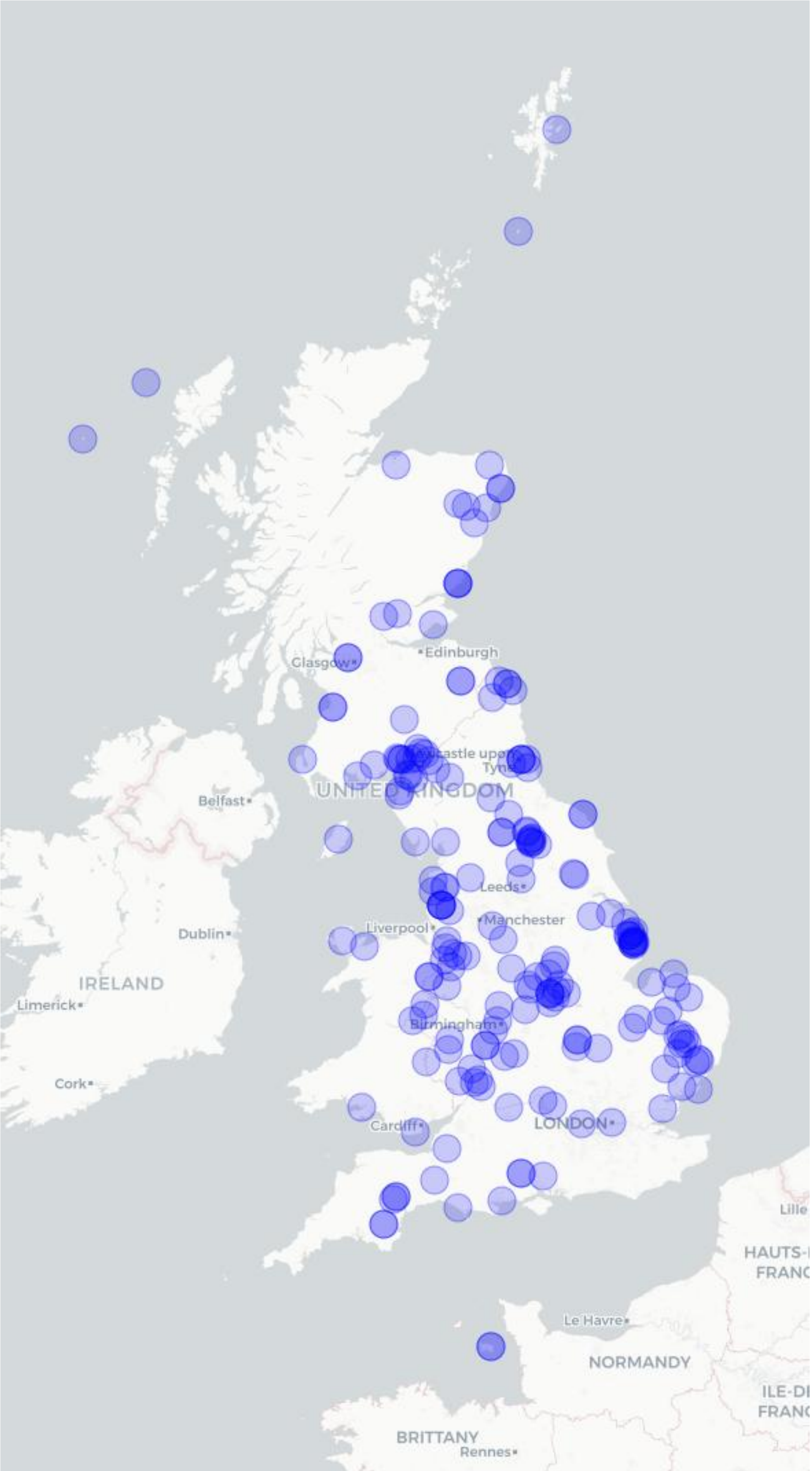

Supplementary Figure 1. Approximate locations of HPAI H5N1 sequenced cases registered in Great Britain in period 03rd December 2020 to 31st May 2022.

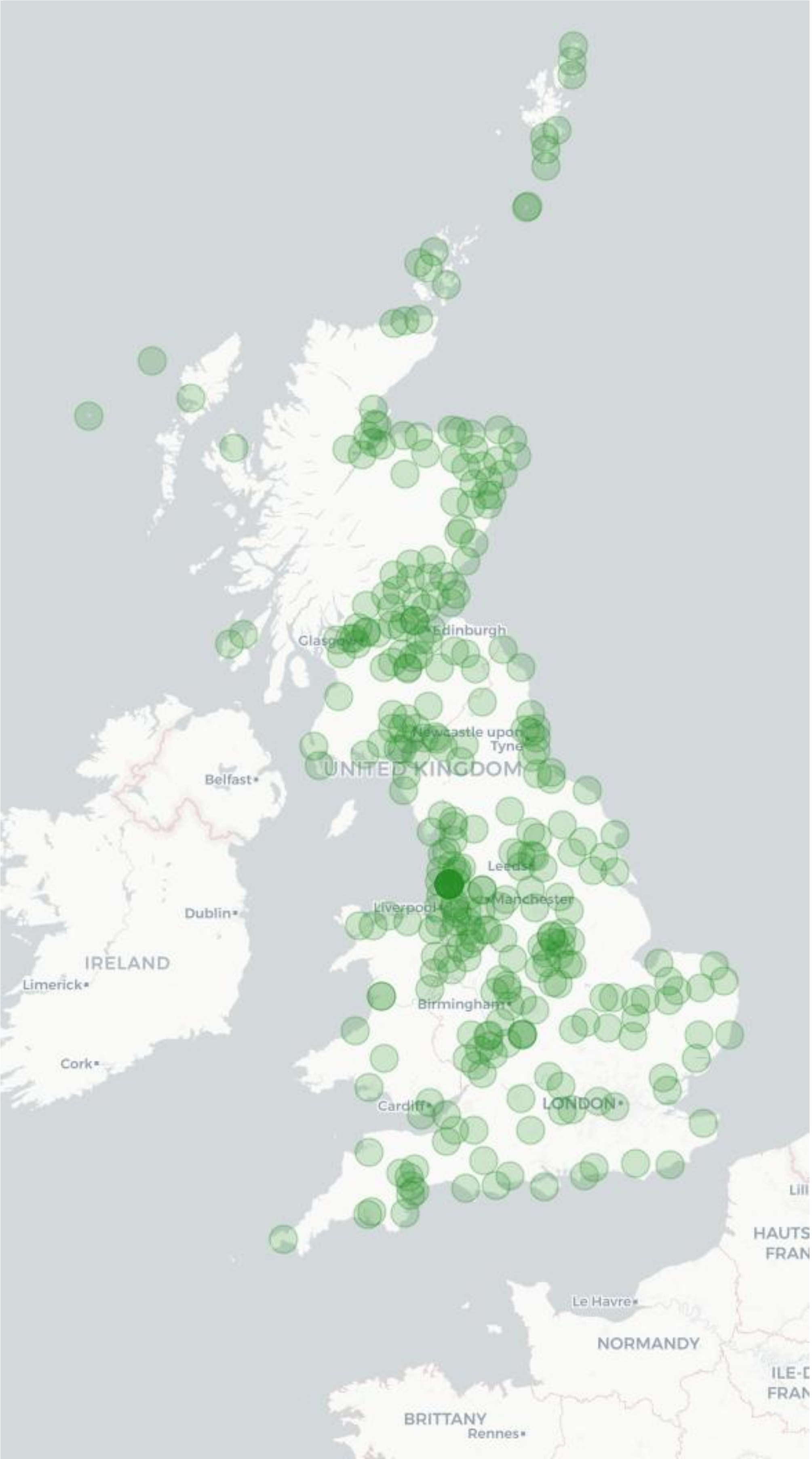

Supplementary Figure 2. Approximate locations of HPAI H5N1 wild bird cases registered in Great Britain in period 03rd December 2020 to 31st May 2022.

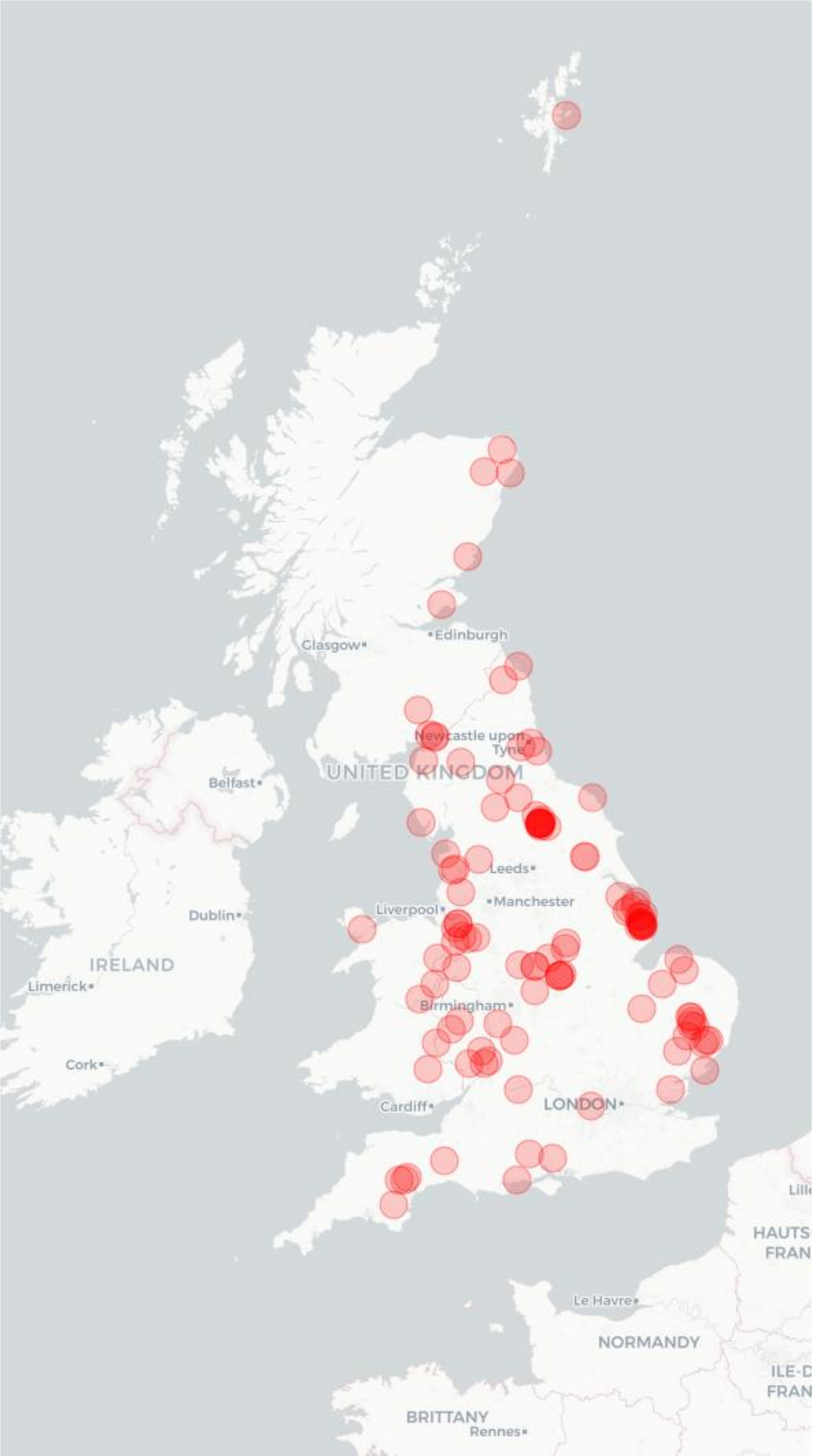

Supplementary Figure 3. Approximate locations of HPAI H5N1 domestic cases (including poultry farms, hobby farms, zoos and similar premises housing captive birds) registered in Great Britain in period 03rd December 2020 to 31st May 2022.

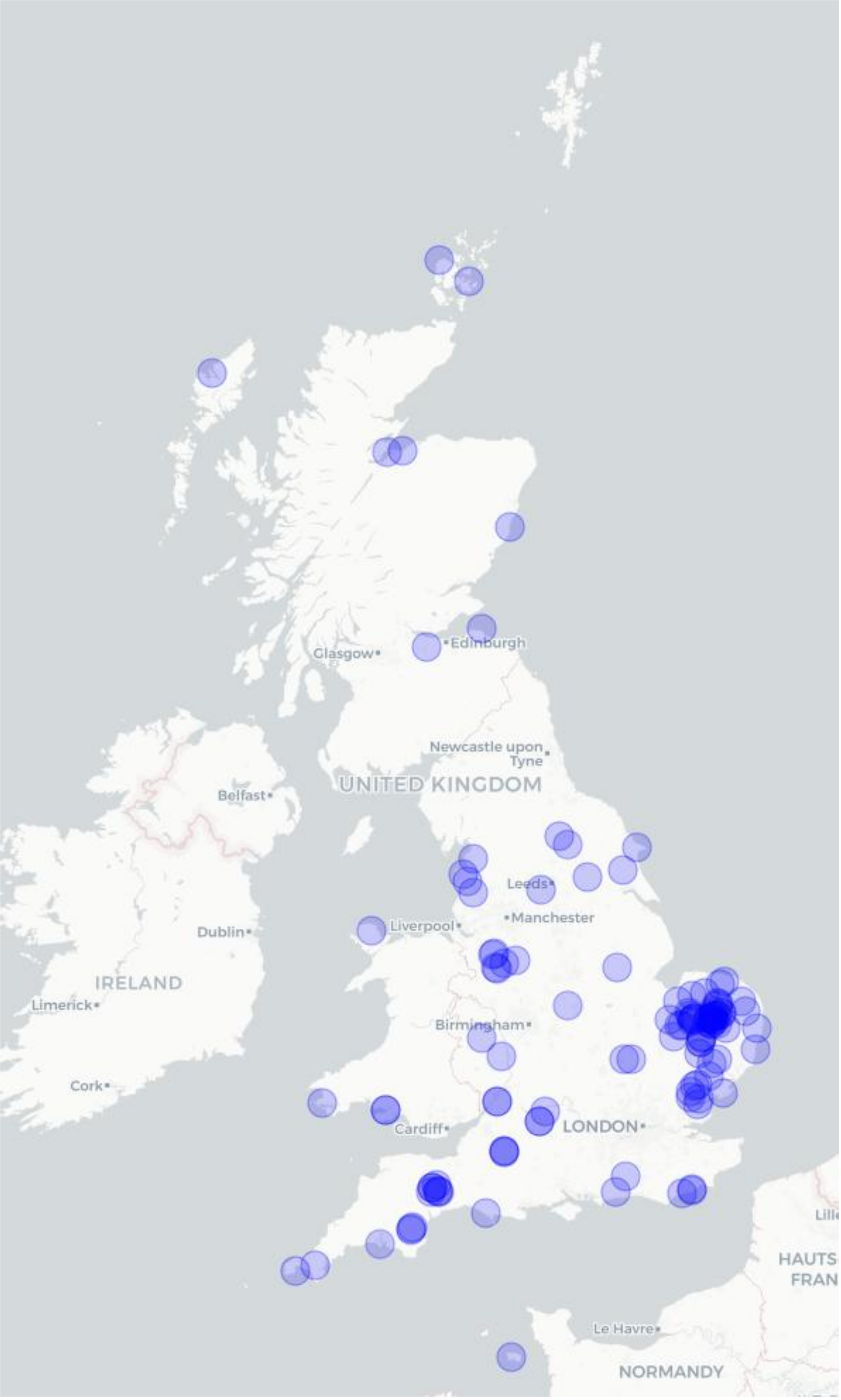

Supplementary Figure 4. Approximate locations of HPAI H5N1 sequenced cases registered in Great Britain in period 01st June 2022 to 19th October 2022.

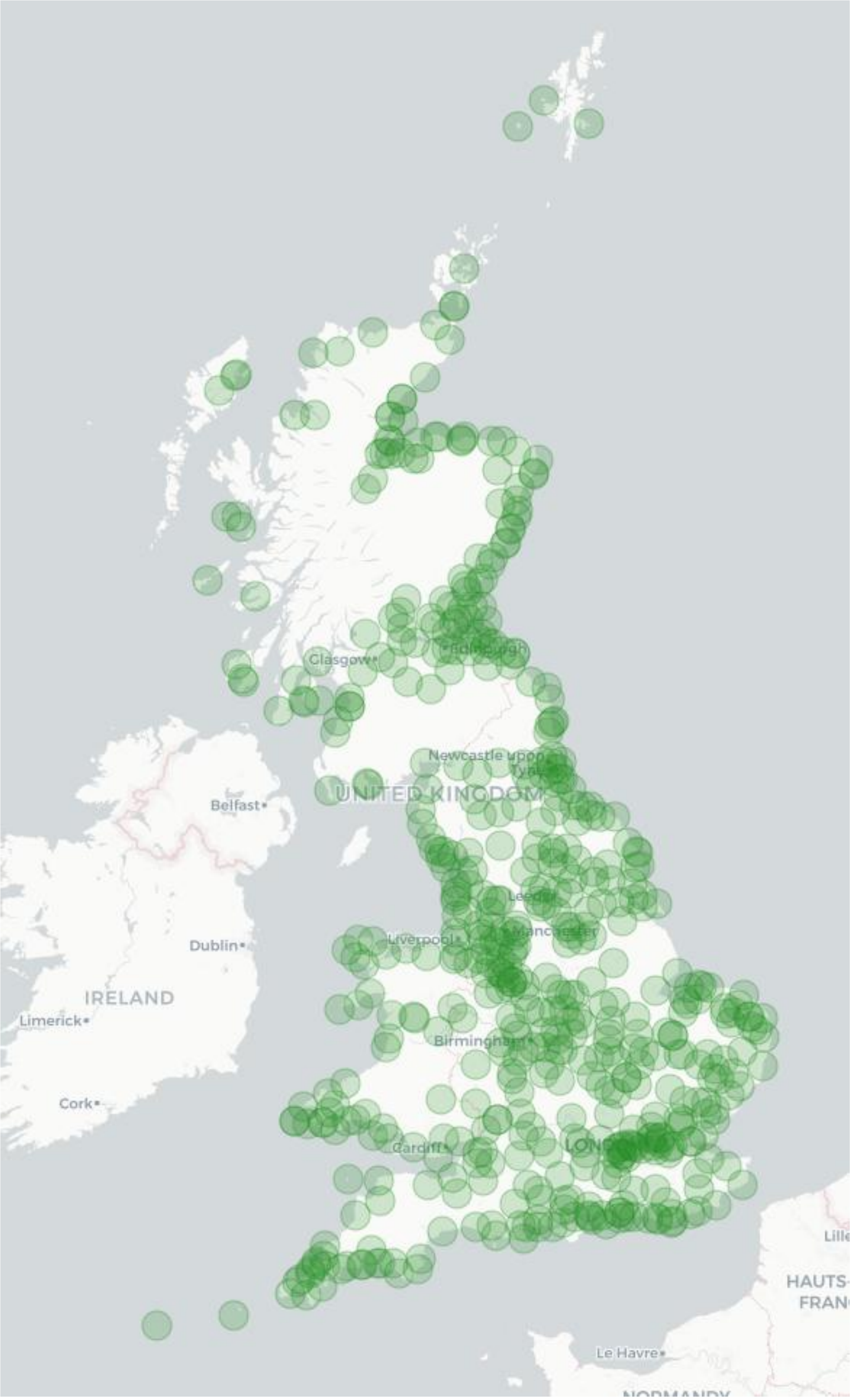

Supplementary Figure 5. Approximate locations of HPAI H5N1 wild bird cases registered in Great Britain in period 01st June 2022 to 19th October 2022.

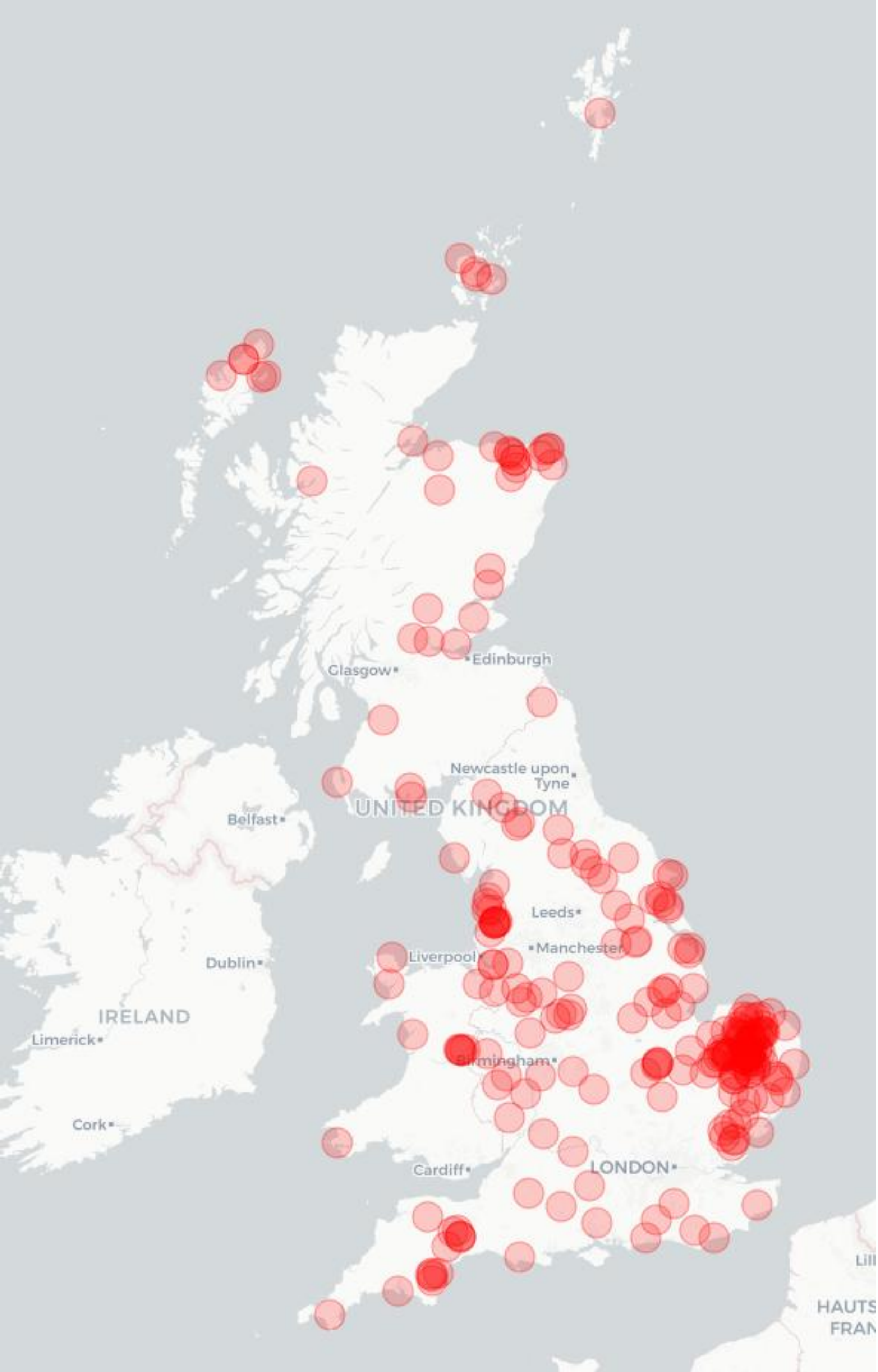

Supplementary Figure 6. Approximate locations of HPAI H5N1 domestic cases (including poultry farms, hobby farms, zoos and similar premises housing captive birds) registered in Great Britain in period 01st June 2022 to 19th October 2022.